\documentclass[12pt,a4paper]{article} 
\usepackage{jheppub,hyperref,mdwlist}
\usepackage{amsmath}
\usepackage{graphicx}
\usepackage{subfigure}
\usepackage{amssymb}
\usepackage{xcolor}
\usepackage{multirow}
\usepackage{cancel}
\usepackage{color}
\usepackage{listings}
\usepackage{xcolor}
\usepackage{float}
\usepackage{slashed}

\usepackage{amsmath}
\usepackage{wrapfig}
\usepackage{cleveref}

\newcommand{\be}{\begin{equation}}
\newcommand{\ee}{\end{equation}}
\newcommand{\beq}{\begin{equation}}
\newcommand{\eeq}{\end{equation}}
\newcommand{\aln}[1]{\begin{align}#1\end{align}}
\newcommand{\bea}{\begin{eqnarray}}
\newcommand{\eea}{\end{eqnarray}}
\newcommand{\besp}{\begin{equation}\begin{split}}
\newcommand{\eesp}{\end{split}\end{equation}}

\newcommand{\nn}{\nonumber}

\newcommand{\Eq}[1]{Eq.~(\ref{#1})}

\newcommand{\Dfbd}{\mathord{\buildrel{\lower3pt\hbox{$\scriptscriptstyle\leftrightarrow$}}\over {D}_{\mu}}}

\def\mE{\mathcal{E}}

\def\mL{\mathcal{L}}
\def\mM{\mathcal{M}}

\def\mP{\mathcal{P}}

\def\0{\textbf{0}}
\def\1{\textbf{1}}
\def\2{\textbf{2}}
\def\3{\textbf{3}}
\def\4{\textbf{4}}
\def\5{\textbf{5}}
\def\6{\textbf{6}}
\def\7{\textbf{7}}
\def\8{\textbf{8}}
\def\9{\textbf{9}}

\def\p{\textbf{p}}

\begin{document}

\title{First-order phase transition and fate of false vacuum remnants}

\author[a]{Kiyoharu Kawana,}
\author[a]{Philip Lu,}
\author[b]{Ke-Pan Xie}

\affiliation[a]{Center for Theoretical Physics, Department of Physics and Astronomy, Seoul National University, Seoul 08826, Korea}
\affiliation[b]{Department of Physics and Astronomy, University of Nebraska, Lincoln, NE 68588, USA}

\emailAdd{kawana@snu.ac.kr}
\emailAdd{philiplu11@gmail.com}
\emailAdd{kepan.xie@unl.edu}

\abstract{
False vacuum remnants in first-order phase transitions in the early Universe can form compact objects which may constitute dark matter. Such remnants form because particles develop large mass gaps between the two phases and become trapped in the old phase. We focus on remnants generated in a class of models with trapped dark sector particles, trace their development, and determine their ultimate fate. Depending on model and phase transition parameters, the evolutionary endpoint of these remnants can be primordial black holes, Fermi-balls, Q-balls, or thermal balls, and they all have the potential to constitute some portion or the whole of dark matter within a broad mass range. Notably, dark sector thermal balls can remain at high temperatures until the present day and are a new compact dark matter candidate which derives its energy from the thermal energy of internal particles instead of their mass or quantum pressure.
}

\maketitle
\flushbottom

\section{Introduction}
Cosmic first-order phase transitions (FOPTs) are ubiquitous in physics beyond the Standard Model (SM). The existence of a FOPT in the early Universe has important phenomenological and observational implications for particle cosmology such as the production of baryon asymmetry due to its non-equilibrium state~\cite{Kuzmin:1985mm,Joyce:1994zt,Joyce:1994fu,Cohen:1993nk,Morrissey:2012db,Huang:2022vkf}, the formation of primordial black holes (PBHs)~\cite{Crawford:1982yz,Hawking:1982ga,Kodama:1982sf,La:1989st,Moss:1994iq,Kusenko:2020pcg,Kawana:2021tde,Huang:2022him,Marfatia:2021hcp,Baker:2021nyl,Baker:2021sno} and more. Since the electroweak (EW) phase transition in the SM is crossover~\cite{Kajantie:1995kf,Rummukainen:1998as,Csikor:1998eu}, new physics effects are necessary to realize a FOPT. Another observable phenomenological consequence is the production of stochastic gravitational waves (GWs)~\cite{Caprini:2015zlo,Caprini:2019egz}, which have drawn attention in recent years with the rapid advances in GW observations~\cite{LIGOScientific:2016aoc,LIGOScientific:2017vwq,LIGOScientific:2018mvr}. The increased sensitivity of GW detectors has enabled tests of new physics models and gravity theories which involve FOPTs~\cite{LIGOScientific:2021sio,LIGOScientific:2021jlr,LIGOScientific:2017adf,NANOGrav:2020bcs,Blasi:2020mfx,Vaskonen:2020lbd,DeLuca:2020agl,Nakai:2020oit,Addazi:2020zcj,Vagnozzi:2020gtf,Benetti:2021uea,Romero:2021kby,Xue:2021gyq}. 

FOPTs can trap particles in the thermal bath, forming macroscopic remnants that might remain up to the present time. This particle trapping mechanism and related phenomena have garnered attention as a new production mechanism of dark matter (DM) and PBHs. The general picture is as follows: When particles have huge mass gaps ($\Delta m\gg T$) between the false and true vacua during a FOPT, most particles do not have sufficient kinetic energy to penetrate the transition wall into the true vacuum (TV) and hence get trapped in the false vacuum (FV). The trapped particles cool and compress into macroscopic objects such as {\it quark nuggets}~\cite{Witten:1984rs,Bai:2018dxf,Zhitnitsky:2002qa,Atreya:2014sca,Gross:2021qgx,Oaknin:2003uv,Lawson:2012zu,Frieman:1990nh}, {\it Fermi-balls}~\cite{Hong:2020est,Marfatia:2021twj} and {\it Q-balls}~\cite{Krylov:2013qe,Huang:2017kzu}. Depending on the model parameters, an initial or secondary collapse into PBHs is possible~\cite{Kawana:2021tde,Huang:2022him,Marfatia:2021hcp,Baker:2021nyl,Baker:2021sno}.\footnote{See Refs.~\cite{Arakawa:2021wgz,Asadi:2021yml} for other baryogenesis and dark matter mechanisms involving particle trapping, and Refs.~\cite{Liu:2021svg,Hashino:2021qoq,Jung:2021mku} for other mechanisms of PBH formation during a FOPT.} Those remnants can not only contribute to DM relics, but can also imprint their presence in the early Universe via decays, evaporation and energy emission (and corresponding entropy production).

A careful study of the post-FOPT evolution of remnants is necessary to understand their effects on the thermal history, which may vary drastically. For example, in the context of PBH formation from FV remnant collapse, Refs.~\cite{Baker:2021nyl,Baker:2021sno} focus on a rapid and direct collapse, whereas Refs.~\cite{Kawana:2021tde,Huang:2022him,Marfatia:2021hcp} focus on a secondary collapse of the non-topological solitons which form at the first stage of FOPT. The former scenario is a non-thermal equilibrium process and thus has to be resolved via numerical simulations, while the latter relies on a pre-existing charge asymmetry (may be related to the baryon asymmetry) such that non-topological solitons can form and allows for an analytical treatment. 

In this paper, we discuss the evolution of FV remnants in another regime where analytical methods can be applied by assuming local thermal equilibrium of trapped particles. While this assumption may not be valid over the entire internal region of the remnant, our results are consistent with the previous numerical studies~\cite{Baker:2021nyl,Baker:2021sno} and supplement our previous work~\cite{Kawana:2021tde} where we have simply assumed that the remnants form non-topological solitons after cooling without any detailed calculations on the intermediate process. In general, the fates of the FV remnants after the FOPT depend on the parameters of the new physics model, and various possibilities have been considered in the literature. In the second part of this paper, we summarize and classify their evolutionary endpoints and address their differences based on our analytical results, although we do not claim an exhaustive presentation of all the possibilities. Throughout this paper, we focus on a class of models which has only SM singlet scalar(s) and fermion(s), and our results have numerous applications to realizations within this simple category.

The organization of this paper is as follows. In Section~\ref{sec:setup}, we explain our setup and calculate the energy gain of the particles from bubble walls. In Section~\ref{sec:initial}, we study the shrinking of FV remnants by assuming local thermal equilibrium of trapped particles, and the validity of this assumption is discussed in Appendix~\ref{sec:validity}. In Section~\ref{sec:fate}, we classify the fates of the remnants according to their model parameters. Concluding remarks are given in Section~\ref{sec:conclusion}.

\section{Setup}\label{sec:setup}

In general, we can consider various early-Universe scenarios in combination with various new physics effects.    
In order to clarify our cosmological setup, we first introduce a typical model and briefly review FOPTs as well as particle trapping mechanism. 
See Refs.~\cite{Megevand:2016lpr,Kobakhidze:2017mru,Ellis:2018mja,Ellis:2020awk,Wang:2020jrd} and references therein for more general aspects of FOPTs.

\subsection{Basic model}

We consider a class of models with a dark sector that can realize a FOPT with particles trapped in the FV. The simplest realization of which is 
\be\label{model}
{\cal L}={\cal L}_{\rm SM}+\frac{1}{2}(\partial_\mu^{}\phi)^2-V(\phi)-\frac{\kappa}{2}|H|^2\phi^2+\bar{\chi}i\slashed{\partial}\chi-y_\chi\phi\bar{\chi}\chi~,
\ee
where $\mL_{\rm SM}$ is the SM Lagrangian with $H$ the Higgs doublet, $\phi$ is a real scalar and $\chi$ is a Dirac fermion. This Lagrangian has a global $U(1)$ symmetry under $\chi\rightarrow e^{i\theta}\chi$, which guarantees the conservation of $(N_\chi-N_{\bar\chi})$. After including finite temperature effects, the potential $V(\phi)$ is modified to the effective thermal potential $V_{\rm eff}(\phi,T)$, which triggers a FOPT by vacuum transition from $\langle \phi\rangle=0$ to $v_\phi(T)$ at a transition temperature $T$ below the critical temperature $T_c$ at which the TV and FV are degenerate.

Defining the fraction of space in the FV as $p(T)\equiv e^{-I(T)}$, then $p(T_c)=1$, and $p(T)\to0$ as $T$ decreases and the FOPT completes. Given the vacuum decay rate $\Gamma(T)$~\cite{Linde:1981zj}, $p(T)$ can be derived~\cite{Guth:1979bh,Guth:1981uk}. {\it Percolation} is defined as the time when TV bubbles form an infinite connected cluster, which happens at $p(T_p)=0.71$~\cite{rintoul1997precise}. In this paper, we choose the characteristic temperature $T_*$ as the epoch when FV remnants are not able to form an infinite connected cluster, which happens approximately at $p(T_*)\approx0.29$~\cite{Lu:2022paj}. The strength of a FOPT is measured by the parameter~\cite{Hindmarsh:2020hop}
\be
\alpha=\left[\Delta V_{\rm eff}^{}(T_*)-\frac{T_*}{4}\frac{\partial \Delta V_{\rm eff}^{}(T)}{\partial T}\Big|_{T_*}\right]\bigg/\rho_R^{}(T_*)~,
\ee
where $\rho_R(T_*)=(\pi^2g_*/30)T_*^4$ is the radiation energy density with $g_*$ the number of relativistic degrees of freedom (DOFs), and 
\be
\Delta V_{\rm eff}(T)=V_{\rm eff}^{}(0,T)-V_{\rm eff}^{}(v_\phi(T),T)~
\ee
is the vacuum pressure. The other relevant parameter is defined by
\be
\beta=-H(T)T\frac{\partial \ln \Gamma (T)}{\partial T}~,
\ee
where $H(T)$ is the Hubble constant. In the following, we represent the scalar field vacuum expectation value (VEV) and Hubble constant at $T=T_*^{}$ as $v_*\equiv v_\phi(T_*)$ and $H_*=H(T_*)$, respectively. 

When an elementary field develops a large mass gap between the FV and TV, only very few particles with sufficient kinetic energy can penetrate into the TV bubbles, while most reflect off the bubble wall and are trapped in the FV. In the model of Eq.~(\ref{model}), the trapping conditions are
\be
M_\chi^*\equiv y_\chi v_*\gg T_*,\quad M_\phi^*\equiv \left(\frac{\partial^2 V_{\rm eff}^{}(\phi,T_*)}{\partial \phi^2}\right)^{1/2}\Big|_{\phi=v_*}\gg T_*^{}~, 
\label{trapping conditions}
\ee
for the fermions $\chi$ and scalar bosons $\phi$, respectively. Typically these conditions are satisfied when a FOPT is strong enough $\alpha\gtrsim 1$, $\beta/H_*\lesssim 100$ because the Universe experiences supercooling, which results in very low $T_*$ relative to the critical temperature $T_c$ of FOPT. There are an abundance of new physics models that can realize the above conditions including classically conformal models~\cite{Iso:2009ss,Iso:2009nw,Iso:2012jn,Kawana:2022fum}.

In this paper, we focus on the subset of models where Eq.~(\ref{trapping conditions}) is satisfied for both $\chi$ and $\phi$. The motivation for assuming trapping of both particles is to enable us to work in the analytical regime where local thermal equilibrium is satisfied in the FV remnant. As we will see, if $y_\chi$ is strong enough to maintain thermal equilibrium, then $\chi\bar\chi\to\phi/\phi\phi$ processes are also efficient. Therefore, if the $\phi$ bosons are free to cross the wall, the trapped fermions will quickly disappear via $\chi\bar\chi$ annihilation and hence no overdensity forms. For the same reason, we also assume that the SM portal coupling $\kappa$ is negligibly small in order to avoid energy loss via rapid decays or annihilations of $\phi,~\phi\phi$ into $HH^*$ or other SM particles, although this could provide for a slow cooling rate (see Appendix~\ref{sec:validity} for further discussion). Although we have Eq.~(\ref{model}) in mind as an example of a FOPT, the following discussions are kept as model-independent as possible so that readers can easily apply our results to more general models. 

\subsection{Energy gain from bubble wall}\label{sec:energy gain}

Now let us consider the energy gain of trapped particles by bubble walls. Consider a spherical shrinking FV remnant. In the vicinity of the bubble wall, we can effectively treat it as a plane moving towards the $-\hat z$ direction with a velocity $-v_w$. By denoting the momentum of a particle in the plasma frame as $p^\mu=(E,p_x,p_y,p_z)$ with $E=\sqrt{m^2+p_x^2+p_y^2+p_z^2}$, the transformed momentum in the wall frame is
\be
p'^\mu=(E',p'_x,p'_y,p'_z)=\left(\frac{E+v_wp_z}{\sqrt{1-v_w^2}},~p_x,~p_y,~\frac{p_z+v_wE}{\sqrt{1-v_w^2}}\right)~.
\ee
After the trapped particle elastically reflects off the wall $p'_z$ flips its sign from $p'_z\to\tilde p'_z=-p'_z$, and the reflected momentum in the plasma frame becomes
\be
\tilde p^\mu=(\tilde E,\tilde p_x,\tilde p_y,\tilde p_z)=\left(\frac{E'+v_wp'_z}{\sqrt{1-v_w^2}},~p'_x,~p'_y,~\frac{-p'_z-v_wE'}{\sqrt{1-v_w^2}}\right)~.
\ee
As a function of initial energy and momentum, $\tilde{E}$ is given by
\be\label{tildeE}
\tilde E=\frac{(1+v_w^2)E+2v_wp_z}{1-v_w^2}~,
\ee
from which the energy gain in a single collision of a particle is calculated as 
\be\label{deltaE}
\delta E=\tilde E-E=2v_w\frac{v_wE+p_z}{1-v_w^2}~.
\ee
Note that $\tilde E$ and $\delta E\to\infty$ for $v_w\to1$ as expected, assuming the particles remain trapped. As a consistency check, we can see that setting $p_z=-v_wE$ yields $\delta E=0$, which means that if a particle is moving with the same velocity as the wall, it does not interact with the wall.

The energy gain per unit area and unit time inside a FV remnant is
\be\label{energy gain}
\mE=\sum_i g_i^{}\int\frac{d^3p}{(2\pi)^3}f_i(p^\mu)\left(\frac{p_z}{E}+v_w\right)\delta E\times \Theta(p_z+v_wE)~,
\ee
where $f_i^{}(p^\mu)$ is the phase space distribution of particle species $i$, $g_i^{}$ is the number of DOFs, and $\Theta(x)$ is the Heaviside step function. It is also easy to check that the pressure ${\cal P}$ by the wall collisions satisfies ${\cal P}=\mE/v_w$. Now the total energy gain is given by
\be\label{DeltaE}
E_{\rm gain}(t)=\int_{t_*}^{t} dt' \mE\times 4\pi r^2(t')=\frac{4\pi}{v_w^{}}\int_{r(t)}^{R_*}dr' r'^2\mE~,
\ee
where $r(t)$ is the radius of the FV remnant at time $t$, and $R_*=r(t_*)$ is its initial radius. The typical (average) size of a remnant at $T_*$ is given by~\cite{Kawana:2021tde,Lu:2022paj}
\aln{\label{Rw}
\langle R_*\rangle\approx \frac{v_w^{}}{\beta}~. 
}
Correspondingly, the number density of FV remnants at $T_*$ is given by~\cite{Kawana:2021tde,Lu:2022paj}
\be\label{remnant density}
n_{\rm rem}^*=p(T_*^{})\left(\frac{4\pi}{3}\langle R_*\rangle^3\right)^{-1}\approx0.07\times \left(\frac{\beta}{v_w^{}}\right)^3~.
\ee
The wall velocity $v_w^{}$ is in general a function of time determined by competing the outward pressure ${\cal P}$ and the inward vacuum energy pressure.

For simplicity, we assume that all the particles inside the remnant (in the FV phase) are massless and have thermal distributions as
\be\label{f}
f_i^{}(p^\mu)=\frac{1}{e^{|\p|/T}\pm 1}~,
\ee
where $+$ is used for fermions and $-$ for bosons, and $T$ is the plasma temperature of the dark sector inside the remnant. We can now perform the integral in Eq.~(\ref{energy gain}), obtaining
\be\label{mP}
\mP=\frac{1}{v_w^{}} \mE=g_d^{}\frac{\pi^2T^4}{90}\frac{(1+v_w^{})^2}{1-v_w^{}}~,
\ee
where $g_d^{}$ is the effective number of DOFs of trapped dark particles. For a stationary wall, $v_w=0$, we see that ${\cal P}$ becomes the usual thermal gas pressure $g_d\pi^2 T^4/90$. In the following, we will discuss the evolution of the remnant profile by using the above analytical results.  

\section{Remnant evolution: initial shrinking}\label{sec:initial}

Neglecting the energy gain from the particle reflection by bubble walls, the energy density inside the remnant simply scales as $\left(R_*/r(t)\right)^3$ due to energy conservation. However, with the inclusion of the reflection energy gain, a more realistic scaling deviates from such a simple scaling as noted in Refs.~\cite{Baker:2021nyl,Baker:2021sno}. In the following, we present an analytical treatment of the evolution of remnant energy and size by assuming that (i) dark sector particles inside the remnant remain in thermal equilibrium and (ii) other energy loss processes such as $\phi\phi\rightarrow HH^*$ are small. The validity of these assumptions are discussed in Appendix~\ref{sec:validity}.

When the remnant is in thermal equilibrium, the energy density is given by 
\be
\rho_d(t)=\frac{\pi^2}{30}g_dT^4(t)~.
\ee
Here, we have put the subscript $d$ for the dark sector to distinguish it from the total radiation energy $\rho(t)$ (including SM particles) which is commonly used in cosmology and set $t=t_*$ at $T=T_*$. The temperature of the remnant will be time-dependent, as denoted by $T(t)$. Applying conservation of energy to the shrinking remnant, we have
\be\label{T(t)}
\frac{4\pi}{3}r^3(t)\rho_d(t)=\frac{4\pi}{3}R_*^3\rho_d^*+E_{\rm gain}(t)-E_{\rm loss}(t)~,
\ee
where $\rho_d^*=\rho_d(t_*)$, $E_{\rm gain}(t)$ is given by Eq.~(\ref{DeltaE}) with Eq.~(\ref{mP}), while the energy loss is modeled by
\be
\label{eq:energyloss}
E_{\rm loss}(t)=\int_0^tdt'\left[\xi_l\rho_d(t')4\pi r^2(t') + \dot{C}\frac{4\pi}{3}r^3(t')\right]~,
\ee
with $\xi_l^{}$ and $\dot{C}$ being the surface and volumetric cooling rates related to the energy loss mechanism, respectively. In the simple model Eq.~(\ref{model}), $\xi_l^{}$ and $\dot{C}$ could describe the effect of the escaping $\chi/\bar\chi$ and $\phi$ particles and decays to SM particles via $\kappa$. For a black body radiation energy loss process~\cite{Witten:1984rs}, $\xi_l=g_l/(4g_d)$ with $g_l$ the number of DOFs of emitted light particles.

By taking the time derivative of \Eq{T(t)}, we obtain
\be\label{T_ODE}
\frac{1}{T(t)}\frac{dT(t)}{dt}=\frac{v_w^{}(t)}{r(t)}\left(\frac{1}{1-v_w^{}(t)}-\frac{v_w^{}(t)}{4}\right)-\frac{1}{r(t)}\frac{3\xi_l}{4}-\frac{\dot{C}}{4\rho_d^{}(t)}~,
\ee
which determines the time evolution of the shrinking remnants for a given wall velocity $v_w^{}(t)$. In particular, for constant wall velocity and $\xi_l^{}=0$ and $\dot C=0$, we have $r(t)=R_*-v_w(t-t_*)$ and \Eq{T_ODE} can be solved as
\be\begin{split}
T(t)\xrightarrow[v_w]{\text{Const.}}&~T_*\left[\frac{R_*}{R_*-v_w(t-t_*)}\right]^{\frac{1}{1-v_w^{}}-\frac{v_w^{}}{4}}~,\\ 
\rho_d(t)
\xrightarrow[v_w]{\text{Const.}}&~\rho_d^*\left[\frac{R_*}{R_*-v_w(t-t_*)}\right]^{\frac{4}{1-v_w^{}}-v_w^{}}~, 
\end{split}\ee
which reduces to $\rho_d(t)\propto r^{-4}$ for small $v_w$, corresponding to an adiabatic compression consistent with Ref.~\cite{Baker:2021nyl}. However, the assumption of constant wall velocity here is oversimplified because the increasing temperature during the shrinking process leads to increasing thermal pressure ${\cal P}$, resulting in a decelerating wall velocity. As mentioned before, we work under the assumption $\xi_l\ll 1$, and in that case it is easy to show that the right-hand-side (RHS) of Eq.~(\ref{T_ODE}) almost vanishes for $v_w(t)\approx 3\xi_l/4$ (the $\dot C$ term is important only when the interaction between trapped particles is extremely small). Based on this fact, the evolution of the remnant can be described in two stages: During the first stage, the remnant is shrinking at a considerable velocity, $v_w(t)\gtrsim\xi_l$, and hence the RHS is dominated by the energy gain term and we can omit the energy loss terms. When $v_w^{}(t)$ becomes $\sim 0$, the $\xi_l^{}$ term starts to dominate and the second stage of evolution, i.e. cooling, begins. The following discussion will cover the first stage of evolution neglecting the $\xi_l$ term, and the cooling will be considered in Section~\ref{sec:fate}.

 The bubble wall initially accelerates after the bubble nucleation due to the vacuum pressure $\Delta V$ until the wall velocity reaches the terminal velocity determined by $\Delta V=\mP$ ~\cite{Ellis:2018mja}. Balancing the thermal and vacuum pressure in Eq.~(\ref{mP}), the wall velocity is given by
\be\label{v(t)}
v_w(t)=4\left(1+\sqrt{1+\frac{8}{3\alpha_d(t)}}
\right)^{-1}-1~,\quad \alpha_d(t)=\frac{\Delta V}{\rho_d(t)}~. 
\ee
Here we assume that the fraction of dark sector particles trapped in the bubble $\sim 1$ and that the temperature in the SM remains the same in both phases. Note that
\be\label{expand}
\alpha_d^*\equiv\alpha_d(t_*)>\frac13,
\ee
is required for the remnant to shrink at $t_*$, and the shrinking stops at $t_1$ when $\alpha_d(t_1)=1/3$. Typically, we have 
$\alpha_d^*\gg \alpha$ because only trapped particles with $g_d\ll 100$ contribute to $\rho_d(t)$. Substituting \Eq{v(t)} into \Eq{T_ODE}, we obtain
\be
\frac{d\alpha_d}{dr}=\frac{3}{r}\alpha_d\left(1+\alpha_d\right)~,
\ee
which can be solved as 
\be\label{alphaf}
\alpha_d^{-1}(r)=\left(\frac{R_*}{r}\right)^3\frac{1+\alpha_d^*}{\alpha_d^*}-1~.
\ee
This result shows that the energy density behaves as $\rho_d(t)\propto r^{-3}$ as a result of the deceleration of the bubble wall during the initial shrinking.

By setting $\alpha_d(r)=1/3$, we can find the terminal radius of initial shrinking as 
\be
R_1\equiv r(t_1)=R_*\left(\frac{1+\alpha_d^*}{4\alpha_d^*}\right)^{1/3}~,
\label{final radius}
\ee
which is always smaller than $R_*$ due to the condition \Eq{expand}. Correspondingly, the temperature of the remnant after this initial shrinking phase is given by
\be\label{terminal temperature}
T_1\equiv T(t_1)=\left(\frac{90\Delta V}{\pi^2g_d}\right)^{1/4}~.
\ee

\section{Evolution of remnant: fate of remnants}\label{sec:fate}
In the previous section, we saw that the initial  shrinking of the remnants terminates when the thermal pressure ${\cal P}$ balances with the vacuum energy pressure $\Delta V$, a general consequence when particle trapping occurs and if energy loss processes from the remnants are not  efficient. On the other hand, the subsequent evolution depends on the model parameters, which control the cooling and collapse conditions.

\subsection{Initial collapse to PBHs}
In Refs.~\cite{Baker:2021nyl,Baker:2021sno}, the possibility of PBH formation during the initial collapse was discussed. A FV remnant can collapse into a BH if its size becomes smaller than its Schwarzschild radius i.e.
\be\label{Schwarzschild criterion}
r<2GE^{(<r)}~,
\ee
where $E^{(<r)}$ is the total energy contained in the remnant and $G$ is the Newtonian constant.
In the present setup, a PBH will form when this condition is satisfied before $r$ reduces to $R_1$. Using \Eq{terminal temperature}, we have 
\be
E^{(<R_1)}=\frac{4\pi}{3}R_1^{3}\left (\rho_d(t_1)+\Delta V\right)=\frac{16\pi}{3}R_1^{3}\Delta V~,
\ee  
which leads to 
\be 
R_1>\sqrt{\frac{3}{32\pi G\Delta V}}~.
\ee
Since the collapse is only possible for large vacuum energy $\alpha_d^*\gg 1$, the final radius Eq.~(\ref{final radius}) can be taken to be the limit $R_1\sim R_*/4^{1/3}\sim v_w\beta^{-1}/4^{1/3}$, and the above condition becomes 
\be\label{PBH condition}
\alpha\gtrsim 0.6\left( \frac{\beta }{v_w H_*^{}}\right)^2~,
\ee
which indicates that a very strong FOPT is necessary for the PBH formation by the initial collapse. This result is consistent with the parameter choice of a large bubble size, $R_* H_*^{}\sim 1$, in Refs.~\cite{Baker:2021nyl,Baker:2021sno}. In such a very strong FOPT, however, particle  trapping may become inefficient~\cite{Baker:2019ndr,Chway:2019kft} because the wall velocity becomes ultra-relativistic, which results in a large boosted energy of particles in the wall frame. Consequently, there is a trade-off between the PBH formation and particle trapping.

\subsection{Thermal ball}

\begin{figure}
\begin{center}
\includegraphics[width=10cm]{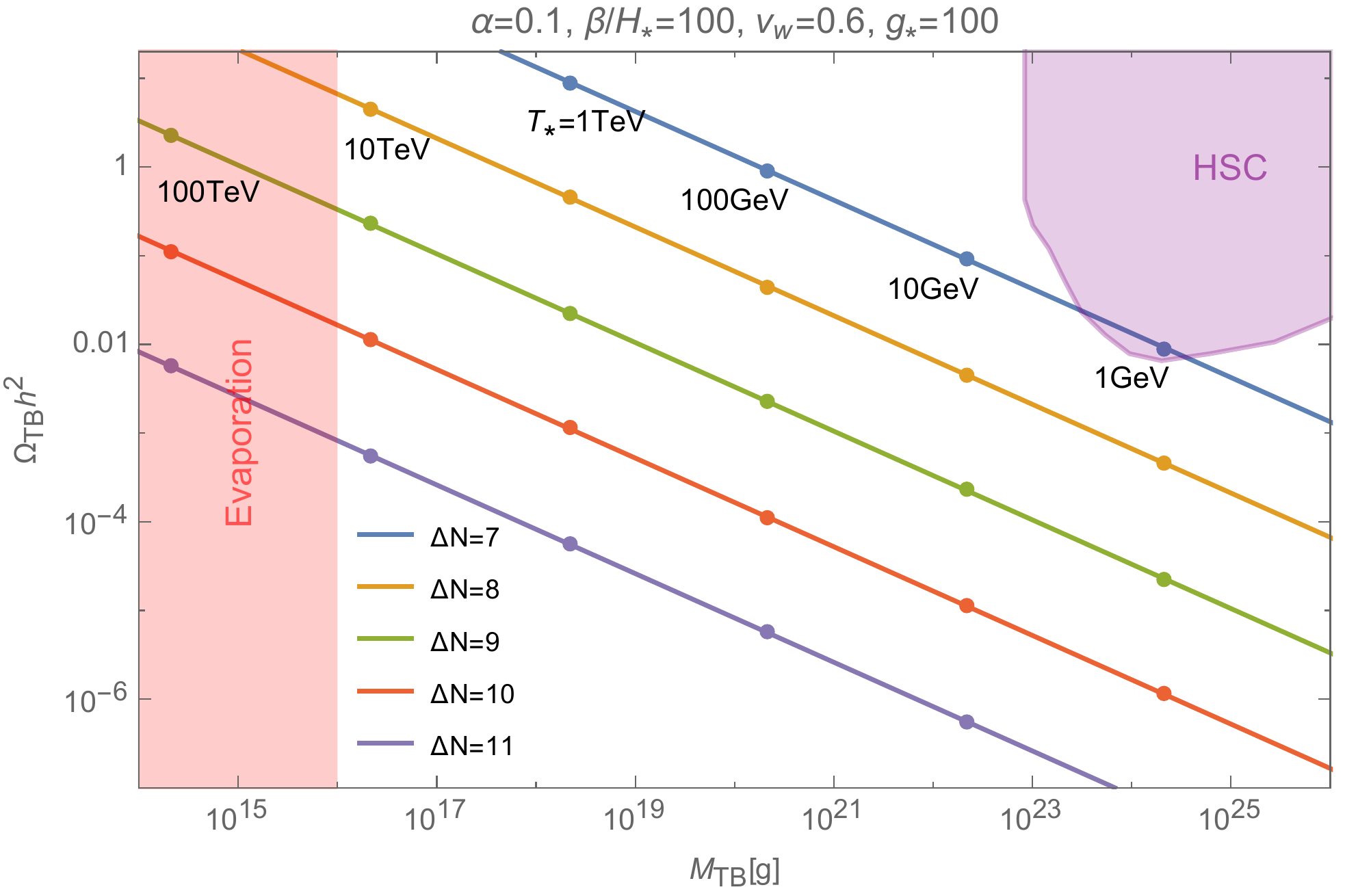}
\caption{Thermal ball/direct collapse PBH abundance. Here different colors correspond to different dilution factors $e^{-3\Delta N}$ and circles mark factors of 10 in the FOPT temperature. Here the purple region corresponds to the gravitational lensing constraint from HSC~\cite{Smyth:2019whb}, while the red region is constrained by Hawking evaporation. Both constraints apply only to PBH remnants.}
\label{fig:abundance}
\end{center}
\end{figure}

If the remnants do not collapse directly to PBHs at the first stage collapse, they will shrink to $R_1$ with the temperature $T_1$. After that, the energy loss term dominates and the remnant starts cooling. Let us first consider an extreme case, $\xi_l\to0$, such that the cooling time scale is greater than the life time of the Universe $\sim 10^{18}$ s. In the model Eq.~(\ref{model}), this can be achieved for extremely small $|\kappa|$ so that decays to SM particles are negligible and $M_\phi^{},~M^{\ast}_{\chi}\gg T_1^{}$ so that energy loss from escaping particles are negligible as well. In this case, the remnants remain in thermal equilibrium with $T \gg T_0\approx2.7 K$ to this day, and they are called ``{\it thermal balls}''~\cite{Gross:2021qgx}. Because $\alpha_d(t_1)=1/3~\leftrightarrow~\rho_d(t_1)=3\Delta V$, the mass of a thermal ball is calculated as 
\be\label{TB mass}\begin{split}
M_{\rm TB}&=\frac{4\pi}{3}R_1^3\times\left(\rho_d(t_1)+\Delta V\right)=\frac{4^2\pi}{3}R_1^3\times \Delta V \\
&\approx 10^{22}~{\rm g}\times v_w^3\left(\frac{100}{g_*}\right)^{1/2}\left(\frac{100~{\rm GeV}}{T_*^{}}\right)^2\left(\frac{100}{\beta/H_*^{}}\right)^3\alpha~,
\end{split}\ee 
where we have used Eq.~(\ref{Rw}). Using Eq.~(\ref{remnant density}), the relic abundance of thermal balls without further dilution is
\bea
\Omega_{\rm TB}h^2&=&h^2\frac{\rho_{\rm rad}}{\rho_{\rm tot}}\frac{\rho_{\rm TB}}{\rho_{\rm rad}}\bigg|_{t=t_0}=0.12\times \left(\frac{\rho_{\rm rad}}{\rho_{\rm DM}}\right)\bigg|_{t=t_0}^{}\times \left(\frac{g_*T_*^3}{g_0T_0^{3}}\right)^{1/3}\frac{\rho_{\rm TB}}{\rho_{\rm rad}}\bigg|_{t=t_*} \nn\\
&\approx&1.2\times 10^{10}\left(\frac{g_*}{100}\right)^{1/3}\left(\frac{T_*}{100~{\rm GeV}}\right)\alpha~,
\eea
where $T_0=2.73~$K and $g_0^{}=3.9$ are respectively the cosmic temperature and number of DOFs today. This result shows that additional dilution is necessary to be consistent with the current DM abundance. Such a dilution can be easily achieved in various new physics models such as thermal inflation~\cite{Lyth:1995hj,Lyth:1995ka,Asaka:1999xd} or early matter era~\cite{Scherrer:1984fd,Berlin:2016vnh,Berlin:2016gtr,Cosme:2020mck}. Note that PBHs formed by initial collapse have the same mass distribution as thermal balls because its mass $M_{\rm PBH}^{}$ is also given by Eq.~(\ref{TB mass}) due to energy conservation. In Fig.~\ref{fig:abundance}, we show $M_{\rm TB}^{}$ vs $\Omega_{\rm TB}h^2$ for varying FOPT temperature and dilution factors, $e^{-3\Delta N}$ with $\Delta N$ the number of additional e-foldings after the FOPT.

Due to their weak interactions with SM particles and relatively diffuse compared to compact remnants such as PBH, if the thermal balls are formed before BBN, constraints on their abundance would be significantly weaker than analogous PBH constraints.

\subsection{Cooling and formation of non-topological solitons}\label{sec:cooling}

When the cooling is sufficiently efficient, the remnants keep shrinking gradually due to the energy loss. Let us first consider the case where surface cooling is dominant. We solve \Eq{T_ODE} for a small $v_w(t) \ll 1$ and surface cooling term $\xi_l^{}\ll 1$, i.e. 
\be\label{eloss}
\frac{dT(t)}{dt}\approx\frac{T(t)}{r(t)}\left(v_w^{}(t)-\frac{3\xi_l}{4}\right)~.
\ee
Thus the cooling term is important when $v_w(t)$ decelerates to $3\xi_l/4$, which happens approximately at $t_1$ as derived in Section~\ref{sec:initial}. The temperature $T(t)$ is related to the velocity via $\mP=\Delta V$ from \Eq{mP}, which under the small $v_w$ expansion is
\be\label{mP_2}
g_d\frac{\pi^2T^4(t)}{90}\left(1+3v_w(t)\right)=\Delta V~,
\ee
Combining above two equations, we find that in the cooling stage the remnant remains at temperature $T_1$ with a constant bubble velocity $3\xi_l/4$. For $t>t_1$, if $v_w^{}<3\xi_l/4$ then $T$ drops and then \Eq{mP_2} cannot hold. Therefore the remnant should remain in a constant temperature and shrink in a uniform velocity. Physically, this can be understood as follows: as the remnant cools, the thermal pressure drops and the inward vacuum pressure condenses the remnant, releasing latent heat and restoring equilibrium. The would-be disappearing time of the remnant is then $t_2=t_1+R_1/(3\xi_l/4)$.

Next we consider volumetric cooling, in particular the case where $\chi/\bar{\chi}$ particles are trapped but $\phi$ scalars are not trapped within the remnant and can instead freely flow out. In this scenario, $\chi\bar{\chi} \rightarrow \phi$ processes can be the dominant cooling mechanism. If the mean free path of $\phi$ is smaller than the bubble size, $\ell_\phi< R_1$, then a steady blackbody radiation is emitted from the surface.  
For one scalar degree of freedom, the remnant will shrink with constant wall velocity $1/4g_f < 1/\sqrt{3}$ below the sound speed. 
This should result in something similar to a deflagration solution~\cite{Espinosa:2010hh}. 
On the other hand, if we have $\ell_\phi \gtrsim R_w$, there will be volumetric emission throughout the entire remnant. The volumetric cooling rate $\dot{C}$ would then depend on the density (temperature) and the annihilation, inverse decay, and scattering ($\chi\rightarrow \chi + \phi$, etc.) cross-sections. We have to solve
\aln{\frac{dT(t)}{dt}\approx\frac{T(t)}{r(t)}\left(v_w^{}(t)-\frac{r(t)\dot{C}}{4\rho_d^{}(t)}\right)~,\quad g_d\frac{\pi^2T^4(t)}{90}\left(1+3v_w(t)\right)=\Delta V~.
}
When $r(t)\dot{C}/\rho_d^{}(t)\ll 1$, we have approximate solution 
\begin{equation}
\label{eq:volumecooling}
   T(t)\approx T_1^{},\quad v_w^{} \approx \frac{r\dot{C}}{4\rho_d}\quad  \Rightarrow\quad  r(t)\approx R_{1}^{} \exp\left({\frac{-\dot{C} t}{4\rho_d}}\right)~.
\end{equation}
As with the case of surface cooling, we expect that the temperature will remain constant at $T_1^{}$ to maintain approximate pressure equilibrium during shrinking. Since the cooling rate $\dot{C}$ depends on the density and average energy of the particles, which in turn depends on temperature, it would also be approximately constant. Long-lasting thermal remnants could still form for small $\dot{C}$, which requires that the interaction rate between $\phi$ and $\chi$ mediated by the coupling $y_\chi$ is minimal. However, such a weak interaction could invalidate our previous assumption of thermal equilibrium during the initial collapse as discussed in Appendix~\ref{sec:validity}. Additionally, we calculate the surface and volumetric cooling rates in the specific case of escaping high energy particles in Appendix~\ref{sec:escape}. In this case, $\dot{C}$ is indeed small for $M_{\chi,\phi}^{}\gtrsim T_1^{}$.

One possible endpoint of cooling is the complete evaporation of the thermal ball remnant. However, the remnant cannot disappear if the outward quantum pressure of the particles can balance the inward vacuum pressure, which results in the formation of non-topological solitons, including the fermionic kind called Fermi-balls~\cite{Hong:2020est} or the bosonic kind called Q-balls~\cite{Friedberg:1976me}. 
Their mass profiles are given by
\be
E_{\rm FB}=Q_f(12\pi^2U_0)^{1/4},\quad E_{\rm QB}=\frac{4\sqrt{2}\pi}{3}Q_b^{3/4}U_0^{1/4},
\ee
respectively, where $Q_f^{}$ ($Q_b^{}$) is the number of fermions (bosons) inside a soliton. 
Even in the simple model Eq.~(\ref{model}), we can consider a  variety of fates of the remnants:  
\begin{itemize}
\item Never disappearing $\phi$ ($\kappa\sim 0$). 
When the decay or annihilation time scale of $\phi,~\phi\phi\rightarrow HH$ is greater than the life time of the Universe, $\phi$ particles never disappear from the remnant.  
In this case, the cooling is dominated by the blackbody radiation by emitted particles with sufficiently high kinetic energy $T\gtrsim M_\phi^{},M_\chi^{}$. When $Q_f^{}\lesssim(\gtrsim ) Q_b^{3/4}$, the remnant looks like  a Q-ball (Fermi-ball) after a long cooling. 

\item Gradually disappearing $\phi$ ($\kappa >0$). 
When the decay or annihilation time scale of $\phi,~\phi\phi\rightarrow HH$ is less than the life time of the Universe, $\phi$ particles gradually  disappear and only a finite number of fermions can survive. As a result, the remnants become Fermi-balls when $Q_f^{}>0$. 
\item Additionally, these Fermi-balls may experience a secondary collapse into PBH if the Yukawa force mediated by $\phi$ overcomes the degeneracy pressure~\cite{Kawana:2021tde}. 
\end{itemize} 
As long as $\kappa$ is not extremely small, the remnants terminate in the latter two possibilities, and the resultant cosmological predictions of Fermi-balls/PBHs are already well studied in Ref.~\cite{Hong:2020est,Kawana:2021tde}. Of course, other endpoints could appear if we consider more complex models.  
We would like to investigate such possibilities in future publications.  

\section{Conclusion}
\label{sec:conclusion}

We have discussed the evolution of FV remnants from FOPTs in a general class of models involving trapped dark sector fermions and scalars. Although we perform our calculations in the context of a simple model, Eq.~\eqref{model}, our analysis can be generally applied to models with trapped particles. We trace their progression through the initial stages of collapse in which the bubble wall accelerates from the vacuum pressure but subsequently decelerates due to the build-up of trapped fermions and scalars.

We then delineate the possible fates of these FV remnants. In the case of horizon-scale remnants, the phase transition would produce a large overdensity satisfying the Schwarzchild condition and collapse directly to a PBH. For smaller remnants, the outward thermal pressure from the trapped particles eventually balance the vacuum pressure and stop the approaching bubble wall. If the coupling between the dark sector and SM sector is negligible, then the cooling timescale of these thermal balls could be longer than the lifetime of the Universe. In contrast with previous work on transient thermal balls~\cite{Gross:2021qgx}, our slow-cooling thermal remnants could contribute significantly to the present day matter density. Thermal balls are a new and qualitatively different dark matter candidate in that they are compact remnants whose energy comes primarily from the thermal energy of their constituent particles, rather than their mass (PBH, generic macroscopic compact halo object) or quantum pressure (Fermi-ball and Q-ball). However, in our fiducial model, thermal balls are overproduced and require a dilution mechanism.

The cooling rate after the initial collapse phase can be significant if the coupling $\kappa$ to the SM is large or if $\phi$ is not trapped. Annihilation and scattering processes can produce a steady stream of particles which escape the remnant. Depending on the mean free path of these particles, the shrinking remnants undergo either surface cooling from blackbody radiation or a slower volumetric cooling. Complete evaporation of the remnant can be avoided if there is a conserved charge associated with the dark sector particles, in which case the remnants would cool until supported by quantum pressure forming a superposed Fermi-ball and scalar Q-ball. Energy considerations suggest that the Fermi-ball will transfer its energy to the Q-ball at low temperatures. However, a stable Fermi-ball can exist if there is a dark fermion asymmetry, in which case a secondary collapse to PBH is possible.

\section*{Acknowledgements} 
We would like to thank Iason Baldes for useful discussions and comments and Joshua Ng for helpful advice.    
The work of KK and PL is supported by Grant Korea NRF-2019R1C1C1010050, 2019R1A6A1A10073437. KPX is supported by the National Science Foundation under grant number PHY-1820891, and PHY-2112680, and the University of Nebraska Foundation.

\appendix
\section{Validity of thermal equilibrium}\label{sec:validity}

We discuss the conditions for thermal equilibrium in the shrinking remnants. Particles reflected from the bubble walls follow a non-equilibrium distribution, but can thermalize via collisions with other particles. In the following, we explicitly calculate the $\chi$-$\chi$ scattering rate and compare the thermalization time scale with the shrinking time scale. As other processes such as $\chi$-$\phi$ scattering have similar cross sections, the $\chi$-$\chi$ process alone is enough to enable an order of magnitude comparison. The assumption used in the previous sections that there was insignificant energy loss during the initial shrinking phase of the remnants constrains the valid range of interactions between the dark
sector and SM sector.

For simplicity, we use a toy model where the entire population of particles has initial momentum $|\p|\sim\mathcal{O}(T)$ and calculate the timescale for a reflected particle to return to this momentum. Consider an elastic collision between a reflected $\chi$ and a $\chi$ in thermal  equilibrium, whose momenta are respectively given by
\be
p_1^\mu=(T+\delta E,0,0,-T-\delta E)~,\quad p_2^\mu=(T,0,0,T)~,
\ee
where $\delta E\sim 2v_w^{}T/(1-v_w^{})$ is given by \Eq{deltaE}, and the total energy of the reflected particle is $E_1^{}=T(1+v_w^{})/(1-v_w^{})$. The center-of-mass (CM) frame momentum is then
\be
p_{\rm cm}^\mu=p_1^\mu+p_2^\mu=\left(\frac{2}{1-v_w}T,0,0,\frac{-2v_w}{1-v_w}T\right).
\ee
Hence, the velocity of the CM frame and the Mandelstam variable $\hat s$ of the scattering are respectively
\be
v_{\rm cm}=-v_w,\quad \hat s=4T^2\frac{1+v_w}{1-v_w}\equiv 4E_{\rm cm}^2~.
\ee
The elastic collision in the CM frame is represented as  $p_1'+p_2'\to p_3'+p_4'$, where these momenta are parameterized as
\be\begin{split}
&p'^\mu_1=(E_{\rm cm},0,0,-E_{\rm cm})~,\quad p'^\mu_2=(E_{\rm cm},0,0,E_{\rm cm})~,\\
&p'^\mu_3=(E_{\rm cm},0,-E_{\rm cm}\sin\theta,-E_{\rm cm}\cos\theta)~,\quad p'^\mu_4=(E_{\rm cm},0,E_{\rm cm}\sin\theta,E_{\rm cm}\cos\theta)~. 
\end{split}\ee
The energy loss in such a collision in the plasma frame is
\be
E_1^{}-E_3^{}=\gamma_{\rm cm}v_{\rm cm}(p'_{1z}-p'_{3z})=-\frac{v_w\gamma_w}{2E_{\rm cm}}\hat t=\frac{-v_w}{2(1+v_w)}\frac{\hat t}{T}~,
\ee
where we have Lorentz transformed between the CM and plasma frames.

Now we can derive the thermalization rate as\footnote{The calculation follows the logic in Ref.~\cite{Baldes:2021vyz}.}
\be
\Gamma_\chi^{\rm th}=\frac{1}{\delta E}\frac{dE_1}{dt}=\frac{n_\chi}{\delta E}\int_{-\hat s}^{-m_\phi^2}d\hat t\frac{d\sigma}{d\hat t}(E_1-E_3)=\frac{3\zeta_3g_\chi^{}y_\chi^4}{512\pi^3}\left(\frac{1-v_w^{}}{1+v_w^{}}\right)T~,
\label{thermalization rate}
\ee
where we have used
\be
n_\chi^{}=g_\chi^{}\frac{3\zeta_3}{4\pi^2}T^3,\quad \frac{d\sigma}{d\hat t}=\frac{|i\mM|^2}{16\pi \hat s^2}
=\frac{y_\chi^4}{16\pi\hat s^2}~. 
\ee
Then, $\tau_\chi^{\rm th}=1/\Gamma_\chi^{\rm th}$ corresponds to the time scale of thermalization. In order for $\chi$ particles to maintain thermal equilibrium, $\tau_\chi^{\rm th}$ has to be smaller than the shrinking time scale $\tau_w^{}\sim R_w^{}/v_w^{}\sim 1/\beta$, resulting in the condition
\be\label{thermalization condition}
\frac{\tau_w}{\tau_\chi^{\rm th}}\approx1.6\times \left(\frac{g_\chi}{4}\right)\left(\frac{100}{g_*}\right)^{1/2}\left(\frac{100~{\rm GeV}}{T_*}\right)
\left(\frac{100}{\beta/H_*}\right)\left(\frac{y_\chi}{10^{-3}}\right)^4\left(\frac{1-v_w^{}}{1+v_w^{}}\right)\gtrsim1.
\ee
For $T_*^{}\sim 100~$GeV, this can be satisfied when $y_\chi^{}\gtrsim 10^{-3}$ or $\beta/H_*^{}\lesssim 100$. It is also straightforward to check that the $\chi$-$\phi$ scattering has a similar thermalization rate as Eq.~(\ref{thermalization rate}) (only the numerical coefficient changes). Therefore, we can conclude that $\chi$ and $\phi$ can both remain in thermal equilibrium inside the remnant when Eq.~(\ref{thermalization condition}) is satisfied, i.e. for large Yukawa coupling or slow FOPTs. Also we notice that the annihilation of $\chi\bar\chi\to\phi\phi$ is also the same order of the $\chi$-$\chi$ scattering. Therefore, when Eq.~(\ref{thermalization rate}) holds, the $\chi\bar\chi$ annihilation to $\phi\phi$ is also efficient. That is why we require the trapping of $\phi$ as well to ensure the existence of overdensity.

The requirement that there be no efficient energy loss mechanism also puts constraints on the range of portal coupling $\kappa$ in which our results are valid. The annihilation rate $\phi\phi\rightarrow HH$ is given by
\be
\Gamma_{\phi\phi\rightarrow HH}^{}=n_\phi^{}\langle \sigma_{\phi\phi\rightarrow HH}^{}v\rangle=\frac{\zeta_3\kappa^2T}{128\pi^3}~,
\ee
where $n_\phi^{}=\zeta_3T^3/\pi^2$ is the thermal number density of $\phi$. Contrary to thermalization processes, the annihilation time scale $\tau_\phi^{}=\Gamma_{\phi\phi\rightarrow HH}^{-1}$ has to be longer than $\tau_w^{}$ as 
\be
\frac{\tau_\phi^{}}{\tau_w^{}}\approx 0.78\left(\frac{g_*^{}}{100}\right)^{1/2}\left(\frac{10^{-5}}{\kappa}\right)^{2}\left(\frac{\beta/H_*^{}}{100}\right)\left(\frac{T_*^{}}{100~{\rm GeV}}\right)\gtrsim  1~. 
\ee
For typical FOPT parameters, this requires the portal coupling $\kappa$ to be small, which is naturally predicted in some new physics models such as classically conformal models~\cite{Iso:2009ss,Iso:2009nw,Iso:2012jn,Kawana:2022fum}.

\section{Escaping particles}\label{sec:escape}

We estimate the cooling rate due to high energy particles escaping the remnant, and derive $\xi_l$ and $\dot{C}$ for this type of cooling. Here, we consider only $\chi$ particles although the inclusion of $\phi$ requires only a minor modification due to the differences between fermionic and bosonic distributions. In this scenario, the $\phi$ particles are fully trapped, but the $\chi$ particles are partially trapped, $M_{\chi} > T$, such that the barrier is low enough to allow the high energy tail of the Fermi-Dirac distribution, $f(p^\mu)=1/(e^{|\p|/T}+1)$, to escape. There are two regimes: if the timescale for thermalization (to regenerate the high energy particles in the distribution) $\tau_{\rm therm}$ is much shorter than the crossing timescale $\tau_{\rm cross}\sim r^2/\ell_\chi$, then we effectively have surface cooling as particles in the high energy tail continuously stream out from the surface. If on the other hand, $\tau_{\rm therm} \gg \tau_{\rm cross}$, then the population of high energy particles in the entire remnant will stream out before being replenished in a time $\tau_{\rm therm}$. In this case, we have a slower volumetric cooling.

We first estimate the energy density carried by the tail of Fermi-Dirac distribution capable of overcoming the mass gap. In the limit where $M_\chi \gg T$,
\begin{equation}
    \rho(E>M_\chi)=\frac{g_\chi}{2\pi^2}\int_{M_\chi}^{\infty} \frac{E^3}{e^{E/T}+1} dE \sim \frac{g_\chi T M_\chi^3}{2\pi^2}e^{-M_\chi /T}~,
\end{equation}
where we have taken only the leading order term in $M_\chi/T$. In the surface cooling limit, $\tau_{\rm therm} \ll \tau_{\rm cross}$, $dE/dt = (1/\pi) g_\chi \rho(E>M_\chi) 4\pi r^2$ where the $1/\pi$ factor comes from angular considerations. We can then identify
\begin{equation}
    \xi_l = \frac{\rho(E>M_\chi)}{\pi \rho_d} = -\frac{120 g_\chi}{7\pi^5 g_d} \left(\frac{M_\chi}{T}\right)^3 e^{-M_\chi/T}~.
\end{equation}
In the volumetric cooling limit, we have $dE/dt = \rho(E>M_\chi) 4\pi r^3/(3 \tau_{\rm therm})$, and we identify
\begin{equation}
    \dot{C} = \frac{\rho(E>M_\chi)}{\tau_{\rm therm}} = \frac{g_\chi T M_\chi^3}{2\pi^2 \tau_{\rm therm}} e^{-M_\chi/T}~.
\end{equation}
We see that for both limits, the condition for slow cooling is $M_\chi \gg T$ so that only a small fraction of particles can penetrate the barrier.

\bibliographystyle{JHEP-2-2.bst}
\bibliography{references}

\end{document}